\documentclass[sigconf]{acmart}
\AtBeginDocument{%
  }

\copyrightyear{2025}
\acmYear{2025}
\setcopyright{cc}
\setcctype{by}
\acmConference[CHI '25]{CHI Conference on Human Factors in Computing
Systems}{April 26-May 1, 2025}{Yokohama, Japan}
\acmBooktitle{CHI Conference on Human Factors in Computing Systems (CHI
'25), April 26-May 1, 2025, Yokohama,
Japan}
\acmDOI{10.1145/3706598.3714308}
\acmISBN{979-8-4007-1394-1/25/04}

\usepackage{multirow}
\usepackage{enumitem}
\usepackage{graphicx}
\definecolor{changes}{RGB}{0, 0, 0}
\definecolor{finalChanges}{RGB}{0, 0, 0}

\begin{document}

%\title{Cross, Dwell, or Pinch: Designing Around-device Selection Method On Unmodified Smartwatches}
\title{Cross, Dwell, or Pinch: Designing and Evaluating Around-Device Selection Methods for Unmodified Smartwatches}
\author{Jiwan Kim}
\affiliation{%
  \institution{School of Electrical Engineering, KAIST}
  \city{Daejeon}
  \country{Republic of Korea}}
\email{jiwankim@kaist.ac.kr}

\author{Jiwan Son}
\affiliation{%
  \institution{KAIST}
  \city{Daejeon}
  \country{Republic of Korea}}
\email{miason427@naver.com}

\author{Ian Oakley}
\affiliation{%
  \institution{School of Electrical Engineering, KAIST}
  \city{Daejeon}
  \country{Republic of Korea}}
\email{ian.r.oakley@gmail.com}


%%
%% The abstract is a short summary of the work to be presented in the article.
% 148 words
\begin{abstract}
Smartwatches offer powerful features, but their small touchscreens limit the expressiveness of the input that can be achieved. To address this issue, we present, and open-source, the first sonar-based around-device input on an unmodified consumer smartwatch. We achieve this using a fine-grained, one-dimensional sonar-based finger-tracking system. In addition, we use this system to investigate the fundamental issue of how to trigger selections during around-device smartwatch input through two studies. The first examines the methods of double-crossing, dwell, and finger tap in a binary task, while the second considers a subset of these designs in a multi-target task and in the presence and absence of haptic feedback. Results showed double-crossing was optimal for binary tasks, while dwell excelled in multi-target scenarios, and haptic feedback enhanced comfort but not performance. These findings offer design insights for future around-device smartwatch interfaces that can be directly deployed on today’s consumer hardware. 
\end{abstract}

\begin{CCSXML}
<ccs2012>
   <concept>
       <concept_id>10003120.10003121.10003128.10011754</concept_id>
       <concept_desc>Human-centered computing~Pointing</concept_desc>
       <concept_significance>500</concept_significance>
       </concept>
   <concept>
       <concept_id>10003120.10003123.10011759</concept_id>
       <concept_desc>Human-centered computing~Empirical studies in interaction design</concept_desc>
       <concept_significance>300</concept_significance>
       </concept>
   <concept>
       <concept_id>10003120.10003138.10003141.10010898</concept_id>
       <concept_desc>Human-centered computing~Mobile devices</concept_desc>
       <concept_significance>300</concept_significance>
       </concept>
 </ccs2012>
\end{CCSXML}

\ccsdesc[500]{Human-centered computing~Pointing}
\ccsdesc[300]{Human-centered computing~Empirical studies in interaction design}
\ccsdesc[300]{Human-centered computing~Mobile devices}

\keywords{Around-device sensing, Selection method, Smartwatch, Sonar}
\begin{teaserfigure}
  \includegraphics[width=\textwidth]{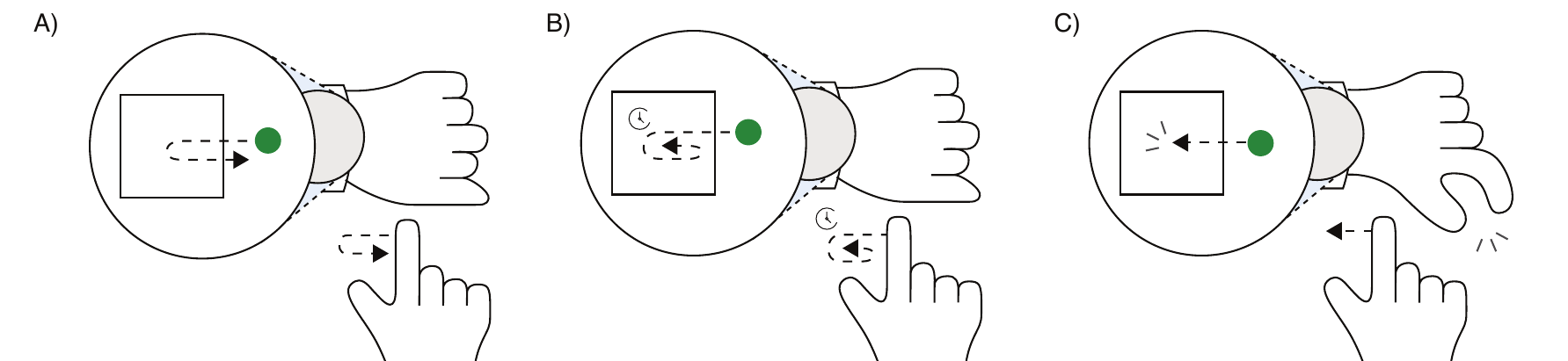}
  \caption{Three target selection methods for around-device interaction on unmodified smartwatches. Double-crossing A) activates selection by moving the cursor onto and then off a target via the same edge. Dwell B) selection is achieved by maintaining the cursor over the target for 500ms. Pinching C) involves making a thumb-to-index finger tap with the hand wearing the watch.}
  \Description{Three target selection method designs for around-device interaction on unmodified smartwatches. Figure 1-A illustrates the double-crossing method, which activates selection by moving the cursor into and then out of a target in the same direction. Figure 1-B shows the dwelling method, which activates selection by maintaining the cursor within the target boundary for more than 500ms. Lastly, figure 1-C shows the pinching method, which detects a pinch gesture using a built-in motion sensor to trigger target selection.}
  \label{fig:teaser}
\end{teaserfigure}


\maketitle

\begin{figure*}[t!]
\centering
   \includegraphics[width=\textwidth]{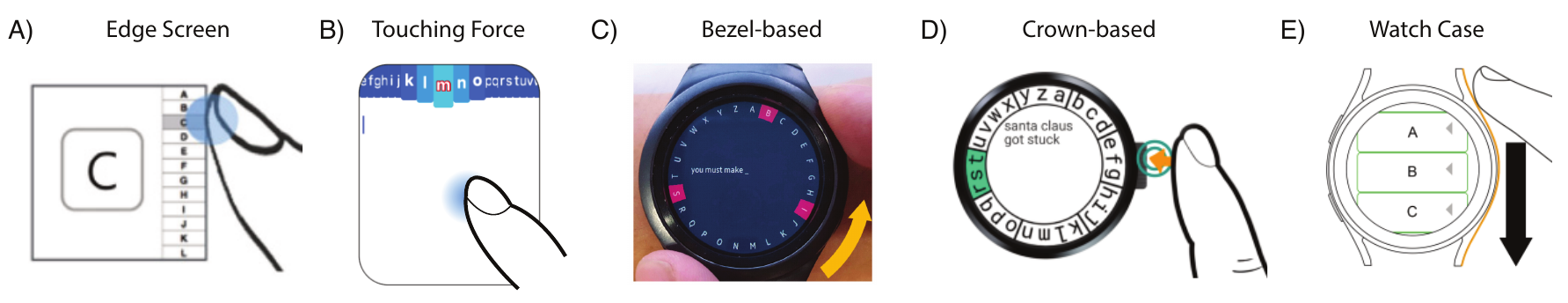}
   \hfil
\caption{\textcolor{changes}{Prior work has proposed various smartwatch systems featuring edge-located uni-dimensional input to avoid occlusion issues during interaction. For example, minimizing the input space by using the edge of the screen~\cite{EdgeInteraction18IJHCSAhn} or the touching force applied to a corner~\cite{1DForceFitts23Ren}, or using physical controls, such as a bezel\textcolor{finalChanges}{~\cite{Compass17CHI}}, crown~\cite{CrownBoard23Rakhmetulla}, or case~\cite{CaseTouch24MUM}}}
\Description{
Figure 2 illustrates prior works that have suggested various interactions on smartwatches with edge-located uni-dimensional layouts to avoid occlusion during the interaction. A) illustrates a user selecting one of the edge-located character arrays, B) shows a user manipulate one-dimensional keyboard using touching force to minimize interaction space. Additionally, C) is bezel-based, D) is crown-based text input and E) utilizes a watch case for uni-dimensional input.
}
\label{fig:stateoftheart}
\end{figure*}

\section{Introduction}
Smartwatches are widely used, compact, and powerful wearable computing devices. They provide many useful features, including notifications, health monitoring, transactions, and communication. However, their small touch screens limit interaction efficiency, as only a few targets can be displayed, and touches often result in the fat-finger problem~\cite{FatFinger05Siek, NanoStylus15UIST}, where the user's finger obscures content. To address these issues, researchers and developers have proposed three main approaches. The first eschews screen-mediated input entirely and focuses on the recognition of discrete sets of single-hand gestures\textcolor{changes}{~\cite{SSI16Reviewers, DigitSpace16Reviewer}}, such as Apple Watch's double tap~\cite{ApplewatchAssitiveTouch}, via modalities including radar~\cite{Headar23Radar}, \textcolor{changes}{camera~\cite{CameraCOTSwatch23Reviewer}}, sonar~\cite{EchoWrist24Lee}, and motion sensing~\cite{Serendipity16Wen, GestureCustomization22Xu}. While this enables one-handed use, it requires learning unprompted hand gestures and lacks the familiarity of manipulating on-screen content. 
The second approach focuses on single-handed cursor or \textcolor{changes}{viewport control} by sensing motion~\cite{Float17Sun} or flexion~\cite{WrisText18Gong} of the wrist \textcolor{changes}{or forearm~\cite{peephole14Reviewer}}. However, these same-hand actions can disrupt views of the watch and induce fatigue. Finally, the third approach assumes two-handed operation of the watch but offsets input from the screen by relying on either \textcolor{changes}{touches to the extreme periphery (Figure~\ref{fig:stateoftheart}, A-B), touches to side-mounted physical input devices (see Figure~\ref{fig:stateoftheart}, C-E), a popular approach prominently exemplified by Apple's digital crown~\cite{AppleCrown} and Samsung's rotating bezel~\cite{samsungBezelGalaxy}, or via} around-device sensing, a term referring to tracking movements of an object (e.g., ring)~\cite{SkinTrack16Zhang, Abracadabra09Harrison} or the hand~\cite{RadarHand24Hajika, AuraSense16Zhou, watchSense17Sridahar} in the region around the watch to control an on-screen pointer~\cite{RadarHand24Hajika, SkinTrack16Zhang, watchSense17Sridahar, AuraSense16Zhou}. This last approach yields considerable benefits: it enables a cursor control paradigm that avoids occlusion and combines flexible, continuous, and familiar input with visually cued and tightly coupled output. 

\textcolor{changes}{While the longstanding integration of physical controls to support offset input into major smartwatch products attests to the fundamental value of such functionality, the practical development of more nuanced and advanced systems based on around-device input is currently barred by two practical problems.} The first is technical: no solutions exist for commercially available smartwatches. Instead, prior research relied on techniques that range from the discreet but unavailable (e.g., integrating a multi-microphone sonar array~\cite{FingerIO16Nandakumar, SoundTrack17Zhang} or high-frequency radar~\cite{RadarHand24Hajika}) through to the impractical in terms of either the sensor used (e.g., a forearm-mounted depth camera~\cite{watchSense17Sridahar}) \textcolor{changes}{or the requirement to augment the object or finger being tracked (e.g., with a magnet~\cite{Abracadabra09Harrison, AuraRing20IMWUT, DigitSpace16Reviewer}), a potentially unrealistic burden for users.} Given that input effectiveness varies with sensing system fidelity, we identify a need to develop solutions that enable around-device input on standard smartwatches. We argue this will advance practical, rather than speculative, research in this area. The second problem is conceptual: while there is consensus on the utility of offset input to control a cursor's position, there is less agreement on the mechanisms for users to trigger target selection in around-device scenarios. Prior research has used techniques ranging from an in-air hand gesture~\cite{esteves2020comparing, PinchClickDwell21Mutasim, PushTap22Dube} (e.g., pinch or in-air tap) or dwell~\cite{esteves2020comparing, Jude14}, a simple technique that requires only additional time over the target (of between 300 ms~\cite{PinchClickDwell21Mutasim} to 800 ms~\cite{PushTap22Dube}), through to more sophisticated systems relying on gaze tracking~\cite{PinchClickDwell21Mutasim, paulus2021usability}, external gesture recognition sensors (e.g., Leap Motion Controller)~\cite{Jude14, PushTap22Dube}, or physical buttons on dedicated input devices~\cite{hansen2003command, esteves2020comparing}. These latter techniques cannot, in practice, be deployed on a watch. In contrast to other device form factors (e.g., phone~\cite{PreTouch16Hinckley, PressTilt18Ando}, HMDs~\cite{esteves2020comparing}), we note no literature explicitly designs and compares selection methods for around-device input on smartwatches. 

This paper addresses these two research gaps. We do this by first adapting a smartphone-based fine-grained sonar tracking system~\cite{LLAP16Wang} to the more limited environment of a commercially available single-microphone smartwatch. \textcolor{changes}{We selected sonar due to its reliance on built-in components (standard microphones and speakers) that allow it to be implemented on unmodified consumer devices~\cite{LLAP16Wang, SonarID22Kim, FingerIO16Nandakumar, LipWatch24Zhang}. However, although the potential of this modality for 2D cursor control is well established for the multiple microphones available on smartphones through techniques like orthogonal frequency division multiplexing~\cite{FingerIO16Nandakumar}, phase changes in continuous waves~\cite{LLAP16Wang}, and impulse response estimation~\cite{VSkin18Sun}, challenges in adapting these techniques to the single microphones available in smartwatches has meant existing work has been limited to exploring specific applications such as finger identification~\cite{SonarID22Kim}, user authentication~\cite{SonarAuth23Kim}, and silent speech recognition~\cite{LipWatch24Zhang} rather than cursor control. To address this issue and achieve a cursor control system, we were inspired by existing offset-pointing input systems on smartwatches. Such systems, typically based on rotating mechanical components such as dials~\cite{AppleCrown} or bezels~\cite{samsungBezelGalaxy}, implement one-dimensional input to achieve practical, meaningful input such as scrolling, menu traversal, and zooming. By targeting this established 1D input scenario, we realized the first high-fidelity sonar-based cursor control system for unmodified smartwatches. Furthermore, we open-source our system~\footnote{https://github.com/witlab-kaist/SonarSelect} to promote future investigations of sonar sensing for around-device input on commercial smartwatches.}

\textcolor{changes}{We then address the second issue, designing viable mechanisms for triggering selections,} by implementing three different methods, each of which avoids screen occlusion and relies on built-in watch sensors. They are: Double-crossing~\cite{nakamura2008double, hakoda2015airflip}, Dwelling~\cite{muller2007dwell, FittsDwellHMD18Hansen, PinchClickDwell21Mutasim}, and Pinching~\cite{PinchClickDwell21Mutasim, GazePinch17Pfeuffer}. We next examine their performance in two studies. First, in a serial binary target selection task and then in a multi-target scenario with and without the support of haptic cues. \textcolor{changes}{Overall, the results show strong user performance and usability, but an interesting trade-off was observed: Double-crossing performs well in binary tasks, while Dwelling yielded benefits in multi-target scenarios. Furthermore, haptic feedback increased comfort ratings. These practical findings and observations can directly contribute to the design of future in-air input techniques on smartwatches.} 
\section{System}
We developed \textit{SonarSelect} on an unmodified Galaxy Watch 6 featuring a built-in speaker, microphone, and 37.3mm display. For sonar sensing, we adapted LLAP~\cite{LLAP16Wang} by configuring it to run at 25Hz \textcolor{changes}{(with a processing latency of less than 15ms) and by} adjusting \textcolor{changes}{hyper-parameters in the Local Extreme Value Detection (LEVD) algorithm to focus on} nearby movements (0-10 cm). Furthermore, we applied a tuned One-Euro filter~\cite{Casiez12} to minimize noise (e.g., cursor jitter). In addition, SonarSelect records IMU data at 100Hz. 

\subsection{System Evaluation}
\begin{figure}[ht]
\centering
   \includegraphics[width=8.5cm]{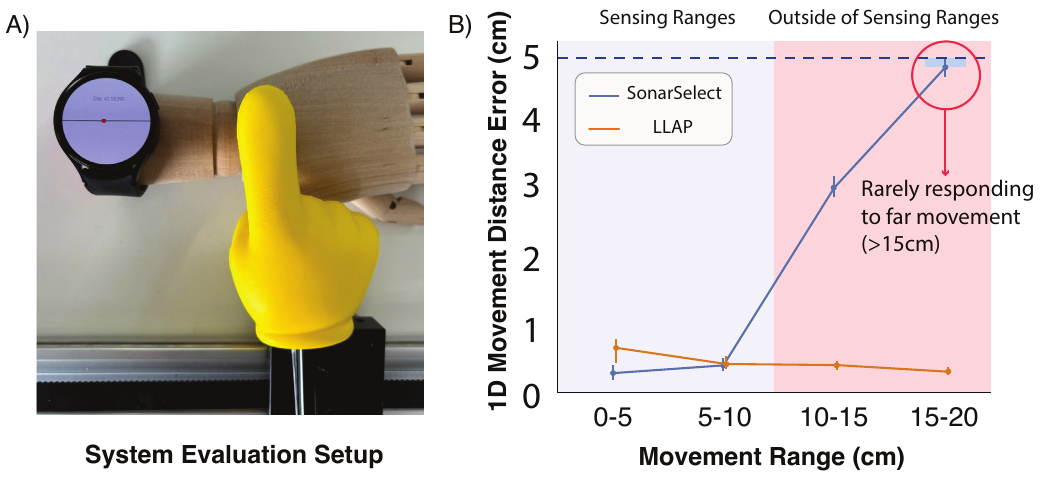}
   \hfil
\caption{\textcolor{changes}{A) SonarSelect evaluation setup: a smartwatch worn on a prop forearm and a prop finger mounted on a linear stage that moves left and right over the back of the prop hand and wrist. B) Mean movement distance error at different movement ranges captured using this setup and compared to a baseline implementation~\cite{LLAP16Wang}. This data shows that SonarSelect shows strong performance for movements within 10cm of the smartwatch.}}
\Description{Figure 3 illustrates the evaluation setup with smartwatches and FingerBot on the left image and measured movement distance error up to different movement ranges on the right images. For the left image, a smartwatch is worn on a prop forearm, and a prop finger is mounted on a linear stage that moves left and right over the back of the prop hand and wrist. For the right image, the Mean movement distance error at different movement ranges was captured using this setup and compared to a baseline implementation. This data shows that SonarSelect shows strong performance for movements within 10cm of the smartwatch.}
\label{fig:system}
\end{figure}

\textcolor{changes}{To verify SonarSelect's performance, we conducted a system evaluation of its sensitivity to object movement at different ranges and speeds using a stepper motor driven linear stage capable of precisely moving a prop finger over a prop hand (see Figure~\ref{fig:system}-A). We also measured the impact of noise in the form of proximate extraneous motion. Consequently, we considered three variables: four \textit{movement ranges} (0-5, 5-10, 10-15, and 15-20 cm), three \textit{movement speeds} (4, 8, and 12 cm/s), and two \textit{noise} conditions (no noise or a person walking around the device at 1-meter distance). For each combination of variables, we captured data from 10 separate 5cm movements, resulting in a total of 240 individual trials. We analyzed the results with a three-way ANOVA, incorporating Greenhouse–Geisser corrections for sphericity violations and following up with Tukey's HSD post-hoc test. There were no interactions, and the only significant main effect was for distance (F (3, 236) = 3895.9, p < 0.001, $\hat{\eta}^2_G$=0.98); all post-hoc tests on this variable showed differences (p<0.001) except between the ranges of 0-5cm and 5-10cm (p=0.12). The mean error in terms of movement range is plotted in Figure~\ref{fig:system}-B. This figure and analysis indicate our sonar system can reliably measure finger displacements, regardless of speed or background noise, and with an accuracy of between 3.45 mm (0-5 cm) and 4.53 mm (5-10 cm) when they occur in the region above the back of the hand; performance beyond that point (i.e., outside of the target sensing range) declines sharply.}

\begin{table*}[t]
\centering
\setlength{\tabcolsep}{2pt}
\Description{Table 1 illustrates existing techniques for target selection using around-device and in-air hand gestures, including Crossing (moving the cursor across a threshold or line), Dwelling (waiting over a region), Pinching (a thumb-to-index tap with the watch hand), and Tapping (e.g., touching the back of the watch hand with the finger of the other hand).}
\caption{\textcolor{changes}{Existing techniques for target selection using around-device \& in-air freehand gestures, including Crossing (moving the cursor across a threshold or line), Dwelling (waiting over a region), Pinching (a thumb-to-index tap with the watch hand), and Tapping (e.g., touching the back of the watch hand with the finger of the other hand)}}
        \label{result_summary_sys}
        \begin{tabular}{c|c|c|c|c}
         \multirow{2}{*}{Method} & \multirow{2}{*}{Illustration} & \multirow{2}{*}{Sensing Technology} & \multirow{2}{*}{Advantages} & \multirow{2}{*}{Disadvantages} \\ &&&& \\
        \hline
        
        \multirow{3}{*}{ Crossing } & 
        \multirow{3}{*}{ \includegraphics[width=1cm]{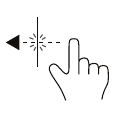} } & 
        \multirow{3}{*}{ Depth camera~\cite{freehandCrossing_TOCHI21} } &  \multirow{3}{*}{ \parbox{4cm}{Quick and enjoyable. Leverage benefits of different techniques~\cite{freehandCrossing_TOCHI21}.} } & 
        \multirow{3}{*}{ \parbox{4cm} {Difficulty in multiple target selection. Physical fatigue during prolonged use~\cite{freehandCrossing_TOCHI21}.} } 
        \\ &&&& \\ &&&& \\ \hline
        
        \multirow{3}{*}{ Dwelling } & 
        \multirow{3}{*}{ \includegraphics[width=1cm]{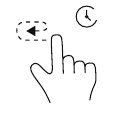} }& 
        \multirow{3}{*}{ \parbox{3cm} {Leap Motion~\cite{PushTap22Dube, PinchClickDwell21Mutasim, Jude14} }} & 
        \multirow{3}{*}{ \parbox{4cm}{Accurate, comfortable~\cite{PushTap22Dube, PinchClickDwell21Mutasim}. Good learnability~\cite{Jude14}. } } & 
        \multirow{3}{*}{ \parbox{4cm} {Slow~\cite{PushTap22Dube, PinchClickDwell21Mutasim, Jude14}. Short dwell time can increase Midas touch problems~\cite{dwelltimeReductionETRA18, EyeHead19Sidenmark}.}}  \\ &&& \\ &&& \\ \hline
        
        \multirow{3}{*}{ Pinching }&
        \multirow{3}{*}{ \includegraphics[width=1cm]{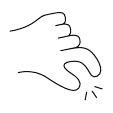}} & 
        \multirow{3}{*}{ \parbox{3cm} {Leap Motion~\cite{PushTap22Dube, PinchClickDwell21Mutasim}, Depth camera~\cite{GazePinch17Pfeuffer, wagner2023fitts} }} &  
        \multirow{3}{*}{ \parbox{4cm}{Comparable throughput with button~\cite{PinchClickDwell21Mutasim, wagner2023fitts}. Benefits when coupled with gaze~\cite{GazePinch17Pfeuffer, PinchClickDwell21Mutasim}. }} & 
        \multirow{3}{*}{ \parbox{4cm} {High recognition errors. Physical fatigue~\cite{PushTap22Dube, PinchClickDwell21Mutasim}.}} \\ &&& \\ &&& \\ \hline
        
        \multirow{3}{*}{ Tapping} &
        \multirow{3}{*}{\includegraphics[width=1cm]{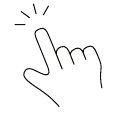}}\ & 
        \multirow{3}{*}{ Leap Motion~\cite{PushTap22Dube, TapLeapMotion15Sensors} } & 
        \multirow{3}{*}{ \parbox{4cm}{Fast, accurate, and minimally demanding~\cite{PushTap22Dube}.} } & 
        \multirow{3}{*}{ \parbox{4cm} { Lower throughput compared to existing controllers (e.g., mouse)~\cite{TapLeapMotion15Sensors}. } } \\ &&& \\ &&& \\ \hline
    \end{tabular}
\label{tab:existingTech}
\end{table*}

\subsection{Selection Triggers}
\label{section:selection_method}
\textcolor{changes}{After achieving robust, reliable finger tracking with SonarSelect, we considered diverse, practical mechanisms for activating target selection during freehand around-device input (see Table~\ref{tab:existingTech}). Implementations for these vary in complexity: Crossing and Dwelling simply involve tracking the position of the cursor relative to targets over time, while Pinching and Tapping can be realized using a watch's built-in motion sensors~\cite{Serendipity16Wen, GestureCustomization22Xu}. \textcolor{finalChanges}{These sensors can detect impact from thumb-to-index pinching or on-skin tapping gestures without requiring contact with the touchscreen, ensuring they do not interfere with sonar-based interactions.} We achieved such a system using a simple, empirically determined threshold on high-pass filtered (15 Hz) smartwatch IMU data~\cite{pinch_detection24}. To validate this system, we captured 800 pinching and tapping gestures from 4 pilot users. Our tuned threshold achieved 95.25\% accuracy (SD 3.75) for pinch detection and a reduced accuracy of 90\% (SD 5.35) for tapping. Further, pinches led to minimal unintended cursor displacement (mean 0.43mm, SD 1.04), while taps were substantially more disruptive (a mean cursor displacement of 3.12mm, SD 5.5). Based on the reduced recognition accuracy and elevated disruption to tracked cursor position, we opted not to include tapping in our studies.} Consequently, we developed three methods to trigger target selection with SonarSelect, each using built-in watch sensors. They were Double-crossing, Dwelling, and Pinching. They are illustrated in Figure~\ref{fig:teaser} and described below:

\textbf{\textit{Double-crossing}}: Double-crossing~\cite{nakamura2008double} involves moving the cursor onto a rectangular target by crossing one of its edges~\cite{Crossing02Accot, freehandCrossing_TOCHI21}. This highlights it as a candidate for selection. Moving the cursor out of the target by crossing the same edge confirms the selection. Leaving the target by crossing another edge cancels the selection. The selection coordinates are captured when the cursor's speed drops to zero as it changes direction.

\textbf{\textit{Dwelling}}: Dwell~\cite{PinchClickDwell21Mutasim, paulus2021usability} involves moving a cursor onto a target and remaining there for a predetermined period of time, after which selection is triggered. We selected a representative 500 ms~\cite{Jude14, GazeGesture15Chatterjee} as the dwell threshold for SonarSelect, following general guidelines for developing interactive applications~\cite{muller2007dwell}. Selection coordinates for dwell are recorded at the end of the dwell time.

\textbf{\textit{Pinching}}: Triggering selection via a pinch of the index and thumb is widely used during interaction on HMDs~\cite{GazePinch17Pfeuffer, wagner2023fitts} and commercial smartwatches~\cite{Serendipity16Wen, GestureCustomization22Xu}. We log selection coordinates when a pinch is detected. 

\section{Study 1: Serial Binary Target Selection}
To evaluate the performance of the three selection triggers, we conducted a one-dimensional Fitt's law study~\cite{Fitts1954information, Fitts92MacKenzie}. The local Institutional Review Board (IRB) approved this study.

\subsection{Participants}
We recruited 18 participants (9 males, mean age of 23 (SD 3.09), mean hand length of 17.75cm (SD 1.82), all right-handed) from the local university. They rated themselves as highly familiar with computers (4.77/5, SD 0.43) and smartphones (5/5) but relatively unfamiliar with smartwatches (2.42/5, SD: 1.42). The study took one hour to complete, and participants were compensated with 15 USD in local currency. Additionally, there was a bonus of 7 USD for the top four participants, based on both input speed and accuracy. 

\begin{figure}[t]
\centering
   \includegraphics[width=8cm]{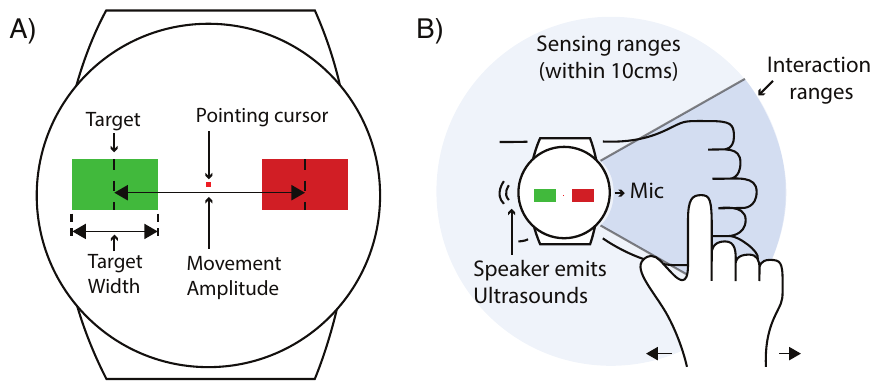}
   \hfil
\caption{Study interface and interaction. Two targets and a cursor are displayed on the smartwatch A). A participant controls the cursor with around-device finger movements B).}
\Description{Figure 4 describes the study interface and interaction. The left image illustrates two targets and a pointing cursor, which are displayed on the smartwatch, and the middle image describes interaction ranges where participants can control the cursor with around-finger movement on the right side of the smartwatch.}
\label{fig:study}
\end{figure}

\subsection{Design}
This study used a one-dimensional Fitts' law serial selection task~\cite{Fitts1954information} with one independent variable, selection trigger, with three levels: Double-crossing, Dwelling, and Pinching. The task is illustrated in Figure~\ref{fig:study}. We used three different target widths (\textit{W}) (3, 6, 9 mm) and two movement amplitudes (\textit{A}) (12 and 15 mm). These values were based on Android accessibility guidelines~\cite{googleTouchTarget} and practical considerations: the use of 3 mm wide targets served as a challenging task, and a maximum amplitude of 15 mm ensured targets always fit on the watch screen. Each trial in the study involved two targets with a given W and A in a series of six consecutive alternating selections. A block of trials was composed of two repetitions of trials with all six possible combinations of W and A delivered in a random order. Each trigger condition in the study featured four trial blocks. The order of trigger conditions was fully balanced among participants, with three completing each of the six possible orders. This resulted in a total of 864 selection tasks per participant: 6 selections by 3 \textit{W} by 2 \textit{A} x 2 repetitions by 4 blocks by 3 triggers.

\begin{figure}[t]
\centering
   \includegraphics[width=7cm]{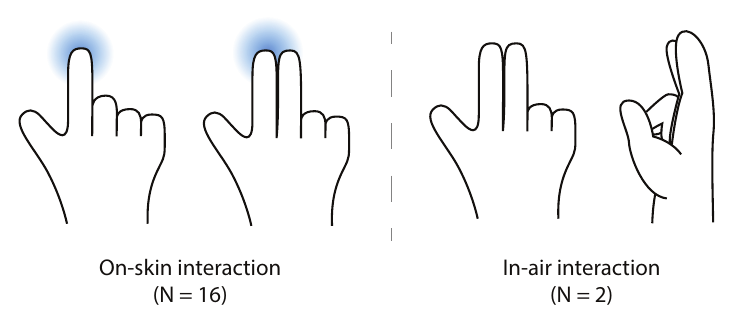}
   \hfil
\caption{The different hand postures adopted by participants in Study 1: Most participants (16/18) opted for on-skin interaction, where they positioned one or two fingers on the back of the hand wearing the watch, while the remaining two participants chose to interact in the air. These choices were not instructed or constrained.}
\Description{Figure 5 describes diverse participants' optimal hand posture during around-device selection. Most participants (16/18) opted for on-skin interaction, where they positioned one or two fingers on the back of the hand wearing the watch, while the remaining two participants chose to interact in the air.}
\label{fig:study1_interaction}
\end{figure}

\begin{table*}[t!]
\centering
% \small
\setlength{\tabcolsep}{5pt}
\Description{Table 1 describes Fitts Law model, quality of fit, mean and standard deviation of throughput, error, and target entries rate of each selection method.}
\caption{The Fitts’ law model, quality of fit ($R^2$), mean (SD) throughput (TP), movement time (MT) error rates (ER), and target re-entries (TRE) of each selection method}
        \label{result_summary}
        \begin{tabular}{c|c|c|c|c|c|c}
         Selection Methods & Fitts’ Law Model & $R^2$  & \textit{TP (bps)}  & \textit{MT (s)} & \textit{ER (\%)}, & \textit{TRE (count/trial)} \\
        \hline
        Double-Crossing  & $MT = 0.065 + 0.475 \times ID_e$ 
        & 0.98  & 2.18 (0.45) & 0.88 (0.26) & 3.29 (1.75) & 0 (0)  \\
        Dwelling & $MT = 0.47 + 0.461 \times ID_e$ 
        & 0.94 & 1.48 (0.24) & 1.34 (0.27) & 0 (0) & 0.12 (0.06)  \\
        Pinching & $MT = 0.021 + 1.234 \times ID_e$
        & 0.90 & 0.92 (0.16) & 1.48 (0.35) & 7.52 (3.25) & 0.12 (0.07) \\
    \end{tabular}
\label{tab:study1_results}
\end{table*}

\begin{figure*}[t!]
\centering
   \includegraphics[width=\textwidth]{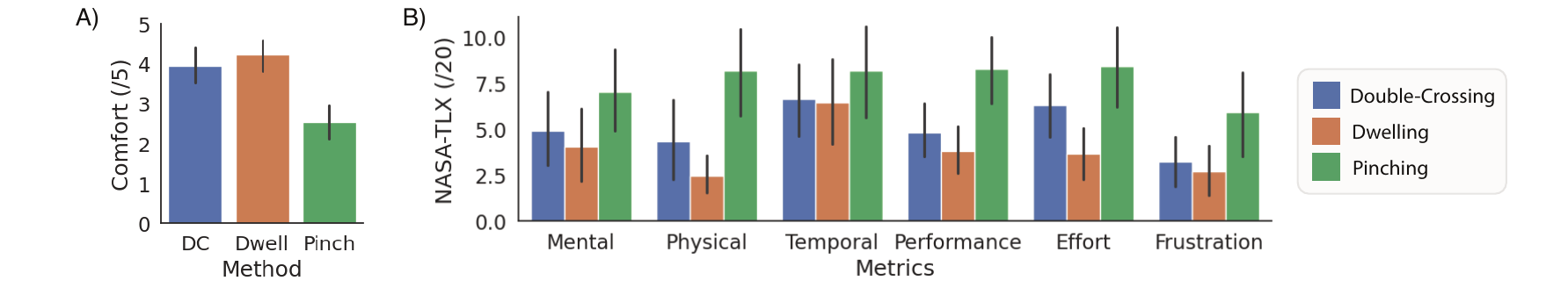}
   \hfil
\caption{Usability questionnaires. Mean perceived comfort A) and workload B) of the three selection triggers.}
\Description{Figure 6 shows the results of usability questionnaires using bar charts. The left image shows mean perceived comfort, and the right image shows the workloads for each selection method. Comfort ratings show high comfort ratings in order of dwelling, double-crossing, and pinching. The pinching selection shows significantly dropped comfort ratings compared to others. Perceived workloads also show similar trends. Overall, it shows high workloads in order of pinching, double-crossing, and dwelling.}
\label{fig:usability}
\end{figure*}

\subsection{Procedure}
The experiment ran in an empty classroom with participants seated at a desk. After signing consent, reading instructions, and having their hand size measured, participants donned the smartwatch on their non-dominant wrist and rested their arms on the desk to prevent fatigue. They were given three minutes to practice with the system and identify a comfortable input pose. Figure~\ref{fig:study1_interaction} shows examples of the hand poses they adopted. Each trial began with participants tapping the screen and positioning their right hand next to the smartwatch’s microphone for a 1-second fixation. Two target boxes were then displayed: one highlighted green for selection and the other red. Participants selected the green target by moving the cursor onto it (by moving their dominant hands/fingers left and right) and performing the appropriate trigger action. After each selection, the highlights switched, and participants selected the newly highlighted target; this process repeated for six consecutive selections. After a trial ended, the next trial immediately began. At the end of each selection method, participants reported their workload using the NASA Task Load Index (NASA-TLX)~\cite{HART1988139}, which enforced a short break. At the end of the study, participants provided demographics and ratings of their comfort.

\subsection{Measures}
Following prior work~\cite{1DForceFitts23Ren, PushTap22Dube}, the primary measures in this study are throughput (\textit{TP})---the rate of information transfer, measured using the effective index of difficulty ($ID_e$) following the revised ISO 9241-9 (9241-411)~\cite{ISO2012}---in bits/second (bps), movement time (\textit{MT})---mean movement time recorded over a sequence of trials---in seconds, error rate (\textit{ER}), and the number of target re-entries (\textit{TRE}). We defined errors as failures to select a target correctly. This occurred when moving off the wrong side of a target during Double-crossing and by tapping thumb-to-index finger together when not over a target for Pinching. However, Dwell lacks a mechanism for erroneous selection of a target~\cite{FittsDwellHMD18Hansen}, so it was expected to record no data on this metric. On the other hand, target re-entries, or the number of times a user moves over (and off) a target before selection, are logged as errors with Double-crossing but recorded separately during Dwelling and Pinching. In addition to these objective metrics, we captured aspects of usability: perceived workload~\cite{HART1988139} and comfort~\cite{Lee21}. 

\subsection{Result}
We recorded 15,552 target selections. The first block in each trigger condition was considered practice and excluded from analysis. Furthermore, 30 selections (0.25\%) were missed due to system failures, meaning we initially retained 11,634 selections. The mean error rate in this set was 4.33\% (SD 3.33). One participant, with an elevated error rate of 16.82\%---more than three SD above the mean---was then excluded as an outlier. Following this procedure, we retained 10,986 selections, with a grand mean movement time of 1.23 seconds (SD 0.14) and error rate of 3.60\% (SD 1.21). All results are presented in ~\autoref{tab:study1_results}.

\subsubsection{\textbf{Throughput and Movement Time}}
We analyzed TP and MT results with one-way repeated measures ANOVAs on the trigger variable, incorporating Greenhouse–Geisser corrections for sphericity violations and following up with Bonferroni-corrected posthoc t-tests to explore pairwise differences. Both TP (F (2, 32) = 155.5, p<0.001, $\hat{\eta}^2_G$=0.77) and MT (F (2, 32) = 83.1, p<0.001, $\hat{\eta}^2_G$=0.62) revealed significant variations with large effect sizes. Post-hoc analysis indicated that Double-crossing yielded the highest TP and lowest MT (all p<0.001), while Dwelling and Pinching differed only in terms of TP (p<0.001). The numerical variations are as stark as the statistical results suggest, with Double-crossing offering improvements of 1.47 and 2.37 times in terms of throughput and 1.52 and 1.68 times in terms of movement time over, respectively, Dwelling and Pinching. 

\subsubsection{\textbf{Error rates and Target re-entries}}
ER was recorded for Double-crossing and Pinching, and TRE was recorded for Dwelling and Pinching; we used paired t-tests to contrast performance between these variables. ER was significantly elevated in Pinching compared to Double-crossing (p<0.001), but no differences were observed in the TRE of Dwelling and Pinching (p=0.61). 

\subsubsection{\textbf{Usability}}
Figure~\ref{fig:usability} shows subjective data. A repeated measures ANOVA on comfort ratings for the trigger technique revealed a significant trend (F (2, 32) = 22.18, p < 0.001, $\hat{\eta}^2_G$=0.42). Posthoc analysis indicated that the only significant pairwise differences resulted from Pinching's reduced perceived comfort of 2.5/5 (SD 0.87) compared to Double-crossing (3.94/5, SD 0.97) and Dwelling (4.24/5, SD: 0.83) (both p < 0.001). Mean overall workload data showed a similar trend with a significant main effect for the trigger technique (F (2, 32) = 12.97, p<0.001, $\hat{\eta}^2_G$=0.23), and posthoc analysis revealing significantly higher demands for Pinching (7.7/20, SD 3.6) compared to Dwelling (3.93/20, SD 2.5) (p < 0.001) and Double-crossing (5.22/20, SD 2.63) (p = 0.037). 

\subsubsection{\textbf{\textcolor{changes}{Post Study Interviews}}}
Comments in post-study interviews were generally positive for Dwelling, but more mixed for Double-crossing. In a representative statement in favor of Dwelling, for example, P16 noted that \textit{``Dwelling selection was intuitive, as I felt like pressing the target"}. However,  while some participants lauded Double-crossing as \textit{``easy, simple, and fast"} (P5, P14) and \textit{``enjoyable"} (P15, P18), others noted that \textit{``Crossing the line didn't intuitively align with confirming target selection"} (P17). Lastly, most participants expressed discomfort with Pinching. It yielded \textit{``many errors"} (11/17) and \textit{``discomfort with using different hands for cursor control and selection"} (5/17).
\section{Study 2: Consecutive Multiple Target Selections}
This study extends Study 1 to examine user performance in a multi-target scenario. It maintains Double-crossing and Dwelling from the first study \textcolor{changes}{but drops the Pinching technique due to its elevated error rates and discomfort ratings.} In addition, to address users' difficulties in associating Double-crossing with triggering target selection, we added haptic feedback to the selection process and examined its impact on performance. The local IRB approved this study.

\subsection{Participants}
We recruited 12 participants (7 males, all right-handed) with a mean age of 22.17 (SD 2.25) and mean hand length of 18.4cm (SD 1.41) \textcolor{finalChanges}{from the local university through an online community.} This study took one hour to complete, and participants were compensated with 15 USD in local currency, with a bonus of 7 USD for the three peak performers.

\subsection{Study Task and Interface}
To select a target size and distance for this study, we examined performance with differently sized and spaced targets in the first study, ultimately selecting 6mm for \textit{W} and 12mm for \textit{A} (see Figure~\ref{fig:study2}-A) as this configuration led to peak performance (see Appendix~\ref{appendix:breakdown_analysis}). To meet our goals of exploring a multi-target scenario, we applied these parameters to display a centered row of three targets at all times. Trials in the study were composed of four selections of these targets. Each trial began with the cursor on the left of the screen and one target highlighted in green. After selecting this target, another target was highlighted. The sequence of the remaining three selections was randomized but structured to ensure there was one repeat selection of the same target, one selection of an adjacent target, and one selection of a non-adjacent target.

\begin{figure}[ht]
\centering
   \includegraphics[width=8cm]{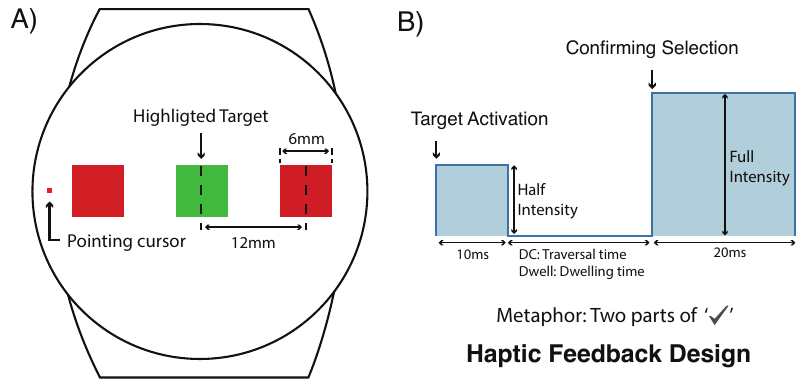}
   \hfil
\caption{Study interface A) and \textcolor{changes}{haptic feedback design for target activation and confirming selection B).}}
\Description{Figure 7 illustrates the study interface and haptic feedback design for target activation and confirming selection in the second study. The left figure shows a watch screen displaying horizontally aligned three targets and a cursor pointer. The right figure shows a half-intensity pulse for target activation and a second longer (20ms) max-intensity pulse for confirming target selection}
\label{fig:study2}
\end{figure}

\subsection{Study Design}
This study featured two independent variables: selection trigger \textcolor{changes}{(\textit{Double-crossing} and \textit{Dwelling})} and haptic feedback. \textcolor{changes}{For this latter variable, we contrasted a baseline of no haptics with a scheme involving two pulses mimicking the short and long strokes of a check mark~\cite{DesignHapticCHI24} using Android one-shot vibration features~\cite{androidHapticFeedback}. Specifically, we used one short (10ms), half-intensity pulse when the cursor moves onto a target and it becomes active (simulating a key press) and a second longer (20ms), max-intensity pulse for confirming target selection (see Figure~\ref{fig:study2}-B).} We fully balanced the order of the \textcolor{changes}{two} trigger conditions and two haptic conditions across the 12 participants. In each condition, each participant completed \textcolor{changes}{480} selections: four selections per trial by six trial repetitions by \textcolor{changes}{five} block repetitions by \textcolor{changes}{two} triggers by two haptic feedback conditions.

\subsection{Procedure}
Study procedures followed the first study: participants sat in an empty classroom, provided consent, read instructions, had their hand size measured, donned the smartwatch and began study trials. \textcolor{changes}{Participants were instructed to control the cursor using on-skin interaction (see Figure~\ref{fig:study1_interaction}), as most participants (16/18) preferred this in Study 1; enforcing a single pose also removes it as a potentially confounding variable.} Participants reported perceived workload and observations after completing each condition and, at the end of the study, provided demographics and comfort ratings for each condition. 

\subsection{Results}
\textcolor{changes}{We recorded 5760 selections and excluded 1152 from the first block of trials as practice and 12 selections (0.21\%) due to system failures, meaning we retained 4596 for analysis.} We analyzed trials by movement time, defined as the time between two consecutive selections, error rates, defined as the proportion of erroneously selected targets, and usability, captured by subjective measures of workload and comfort. 

\subsubsection{Objective Measures}
\begin{figure*}[t!]
\centering
   \includegraphics[width=\textwidth]{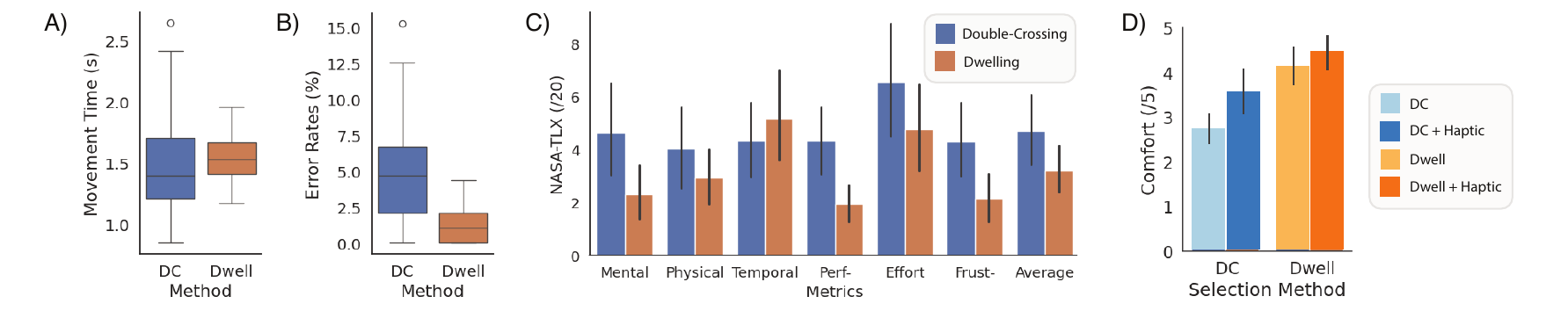}
   \hfil
\caption{User performance and usability data during multiple target selections. Movement time A), error rate B), \textcolor{changes}{and perceived workload C) for the different selection methods}, and comfort rates for the different haptic conditions and selection methods D).}
\Description{Figure 8 illustrates user performance and usability data during multiple target selections. Figures A and B illustrate a box plot of movement time and error rates, and Figure C shows a bar plot of perceived workloads for different selection methods, which illustrates similar movement time between dwelling and double-crossing and high error rates in double-crossing. Lastly, the right figure shows comfort rates for different haptic conditions and selection methods, which shows that all selection methods with haptic feedback provide more comfort than ones without haptic and dwelling, which outperformed double-crossing in comfort ratings.}
\label{fig:study2Result}
\end{figure*}

Objective data are shown in Figure~\ref{fig:study2Result} and were analyzed with two-way repeated measures ANOVAs. \textcolor{changes}{Movement time showed no significant main effects or interactions (F-values in 0.05-3.25, p-values in 0.10-0.83).} Error rates, on the other hand, showed a significant main effect for selection trigger \textcolor{changes}{(F (1, 11) = 9.91, p=0.009, $\hat{\eta}^2_G$=0.29)} but not for haptics or their interaction \textcolor{changes}{(F-values between 0.13-0.28, p-values between 0.60-0.72).} The significant effect indicates that Double-crossing yielded more frequent errors \textcolor{changes}{(5.21\%, SD: 4.14) than dwelling (1.36\%, SD: 1.54).}

\subsubsection{Usability}
Two-way repeated measures ANOVAs on comfort ratings showed significant main effects for \textcolor{changes}{both trigger (F (1, 11) = 22, p<0.001, $\hat{\eta}^2_G$=0.39) and haptic (F (1, 11) = 7, p=0.02, $\hat{\eta}^2_G$=0.13) variables} but no interaction effect \textcolor{changes}{(p=0.21)}. This indicates that the presence of haptics \textcolor{changes}{(4.04/5 vs 3.45/5 in its absence)} and use of \textcolor{changes}{dwelling (4.33/5 vs 3.17/5 with double-crossing) led to significantly greater perceived comfort. Overall workload showed a significant main effect for trigger (F (1, 11) = 21.36, p<0.001, $\hat{\eta}^2_G$=0.07) with a higher mean overall workload for double-crossing compared to dwelling (4.02/5 vs 2.75/5)} - see Figure~\ref{fig:study2Result}-C for all NASA-TLX metrics. However, there was no difference due to the haptic variable and no interaction effect \textcolor{changes}{(F-values between 2.15-3.61, p-values between 0.08-0.16).}

\section{Discussion}
These studies highlight trade-offs in selection trigger design during around-device input. We suggest that the most appropriate selection trigger will depend on both the complexity of the task being supported and user preference.

\textbf{Double-crossing} enabled fluid and rapid performance in binary tasks, \textcolor{changes}{but this advantage was ameliorated} in sequential multi-target tasks because the repeated selections required frequent changes in direction. However, \textcolor{changes}{while Double-crossing accrued more errors and reduced comfort ratings compared to Dwelling, participants also reported it was \textit{``enjoyable"} (P15, P18) and \textit{``game-like"} (P19, P28), comments that complement prior observations of hands-free crossing interaction~\cite{freehandCrossing_TOCHI21}. This suggests Double-crossing can effectively support common watch tasks involving single operations such as dismissing a notification~\cite{SwipeToDismissAndroid} or selecting one command, mode, or menu-item. Furthermore, it may provide a playful aspect to operating physically inspired interactions such as a flicking-~\cite{scrolling14UIST_Juho} or braking-to-scroll~\cite{FlickAndBrake11CHIEA}.}

\textbf{Dwelling} is slower than Double-crossing in binary tasks but supports equivalently rapid input during sequential multi-target tasks. Additionally, it yields consistently low error rates across different tasks. \textcolor{changes}{Further, users report it to be comfortable and place low demands on their mental and physical resources, confirming prior findings~\cite{PushTap22Dube, PinchClickDwell21Mutasim}. These results suggest that Dwelling is broadly applicable: it can effectively support both complex, sequential multi-target selection tasks and maximally reliable performance in simpler, isolated tasks.}

% Finally, we note that the stability in Dwelling performance data between the two studies suggests that SonarSelect is robust to the various hand poses used in Study 1; enforcing a single pose in Study 2 provided few benefits. This suggests that SonarSelect can be used with various hand poses without unduly impacting performance.

\textbf{Pinching} achieved poor objective performance (e.g., lowest TP, highest MT, ER, and TRE) and subjective ratings (e.g., lowest comfort, highest workload) in binary tasks. \textcolor{changes}{This may be due to the Heisenberg effect~\cite{HeigenBergCHI20Dennis}, a term referring to situations where the act of triggering a selection causes users to fall out of their intended target, resulting in an error. To investigate if optimized implementations might alleviate this problem, we explored a simple stability-enhancement technique: when we detect a pinch, we offset the trigger location in time to align it with the position of the cursor at the start of the pinch. We applied a range of fixed offsets (40 to 200 ms, in steps of 40), and while the peak performing value (40 ms) corrected 78.5\% of the existing pinch errors (thus improving the error rate by 5.91\%), it also introduced a similar proportion of false positives (6.11\%) in the form of premature trigger events. Consequently, despite Pinching's demonstrated feasibility for triggering single preset actions in commercial products~\cite{samsungMoreThan, ApplewatchAssitiveTouch}, we conclude it is unsuitable for use as a general-purpose selection trigger due to challenges in coordinating bi-manual input and the elevated error rates that arise from these difficulties.}

Despite these variations, our studies comprehensively demonstrate that sonar-based selection is viable on commercial smartwatches. Grand mean movement times of 1.23s and \textcolor{changes}{1.51s} and error rates of 3.5\% and \textcolor{changes}{3.29\%} in our binary and multi-target selection tasks are compelling and competitive with those reported in the literature. For example, force-taps~\cite{1DForceFitts23Ren}, a recently proposed scheme for one-dimensional target selection on smartwatches involving varying pressure on the watch's touchscreen edge, shows broadly comparable results for throughput (e.g., 2.19 vs 2.21 bps (vs Double-crossing) with 3mm targets), while still occluding screen contents. This comparison suggests that sonar-based selection may be suitable for integration into today's smartwatch products.

\subsection{Future Work}
\textcolor{changes}{SonarSelect, a one-dimensional sonar-based cursor control system, is inspired by existing solutions (e.g., dials~\cite{AppleCrown}, bezels~\cite{samsungBezelGalaxy}) for offset input on smartwatches. Its advantages include not requiring dedicated hardware components, potentially decreasing manufacturing costs, complexity, and device size while increasing reliability and maintaining the same uni-dimensional expressivity of the input. While this paper has demonstrated the fundamental viability of SonarSelect, immediate future work should compare it against the dial and bezel input it was inspired by in genuinely representative watch input tasks such as selecting modes or menu items. This will more comprehensively establish the suitability of the technique for practical deployment. Additionally, future work should explore the feasibility of deploying SonarSelect on smartwatches in terms of integration at an operating system level. Finally, while we have focused on the practical and widely applicable scenario of uni-dimensional cursor control, future work should extend this to explore more sophisticated sonar algorithms and richer around-device input scenarios. For example, Tianhong et al.~\cite{ringAPose_YU_IMWUT24} recently proposed continuous hand-pose estimation using a single microphone and speaker on a custom ring. Although such techniques are arguably too slow for a viable cursor control scenario (peak system latencies of 85ms), they do highlight the potential to develop more expressive systems capable of, for example, recognizing both pose and position or enabling two-dimensional finger tracking. Exploring such systems is a next step for this work.}

\section{Conclusion}
We explored around-device interaction on unmodified smartwatches using sonar-based finger tracking, evaluating selection triggers in binary and multiple target scenarios and revealing that while Double-crossing was fastest for binary tasks, Dwelling was more effective for multiple targets. Haptic feedback also improved usability but did not enhance performance. \textcolor{changes}{Overall, SonarSelect shows compelling user performance and usability, suggesting it may serve as a viable replacement for (or enhancement of) the physical controls currently used for one-dimensional offset input on smartwatches.}

\begin{acks}
Thank you to all the participants. This research was partly supported by the MSIT (Ministry of Science and ICT), Korea, under the ITRC (Information Technology Research Center) support program (IITP-2024-RS-2024-00436398) supervised by the IITP (Institute for Information \& Communications Technology Planning \& Evaluation) and the Basic Science Research Program through the 
National Research Foundation of Korea (NRF) funded by the Ministry of Education (RS-2024-00407732).
\end{acks}

\bibliographystyle{ACM-Reference-Format}
\bibliography{base}

% currently 4999 w/o appendix. Max is 5176 with appendix.
\appendix

\section{Analysis by target widths and movement amplitudes in Study 1}
\label{appendix:breakdown_analysis}
\begin{figure*}[t]
\centering
   \includegraphics[width=\textwidth]{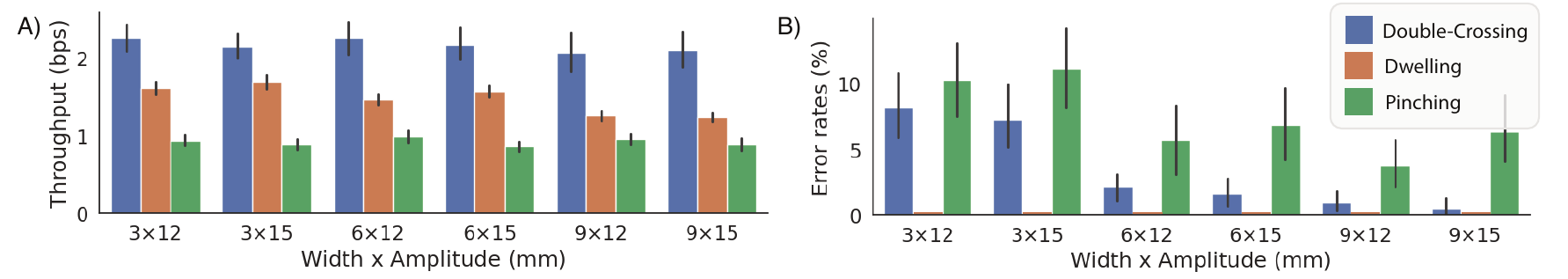}
   \hfil
\caption{Mean throughput A) and error rates B) for three selection methods in study 1 by target widths and movement amplitude.}
\Description{Figure 9 illustrates average user performance data for the three examined selection methods by target widths and movement amplitudes. Figure 8-A shows throughput. The double-crossing selection showed higher throughput compared to the dwelling (1.47 times faster) and pinching (2.36 times faster) selections. Additionally, Figure 8-B shows error rates, which show no error in dwelling. On the other hand, pinching showed the highest error rates.}
\label{fig:AppendixB_result}
\end{figure*}

To determine optimal target sizes for use in Study 2, we analyzed performance in terms of the width (W) and spacing (A) of targets in Study 1 for all three selection triggers. Figure~\ref{fig:AppendixB_result} shows the summary data and Table~\ref{tab:appendixB_stats} the results of the statistical analysis. We ultimately selected W as 6 and A as 12 for use in Study 2 due to this combination yielding high throughput and low error rates. 

\begin{table*}[t]
\centering
% \small
\setlength{\tabcolsep}{5pt}
\caption{Results of a two-way repeated measures ANOVA on target width (W) and movement amplitudes (A) for measured throughput and error rates in study 1.}
\Description{Breakdown analysis with two-way repeated measures ANOVA by target width and movement amplitudes for measured throughput and error rates}
        \label{result_summary_app}
        \begin{tabular}{c | c | c c c c c }
         Metrics & Source & df &  F  & p  & $\hat{\eta}^2_G$ & Post-hoc pairwise comparisons \\
        \hline
        \multirow{3}{*}{\textbf{Throughput}} & \textit{W} & (2, 32)  
        & 28.6  & <0.001 & 0.10 & 3 vs 9 (p<0.001), 6 vs 9 (p<0.001), 3 vs 6 (p=0.10)\\
        & \textit{A} & (1, 16) 
        & 6.6 & 0.02 & 0.004 & 12 vs 15 (p=0.02) \\
        & \textit{W} $\times$ \textit{A} & (2, 32)
        & 0.15 & 0.86 & 0.0002 & - \\

        \hline
        \multirow{3}{*}{\textbf{Error rates}} & \textit{W} & (2, 32)  
        & 31.4  & <0.001 & 0.39 & 3 vs 6 (p<0.001), 3 vs 9 (p<0.001), 6 vs 9 (p=0.16)\\
        & \textit{A} & (1, 16) 
        & 1.11 & 0.31 & 0.004 & - \\
        & \textit{W} $\times$ \textit{A} & (2, 32) & 0.20 & 0.75 & 0.004 & - \\
        
    \end{tabular}
\label{tab:appendixB_stats}
\end{table*}

\end{document}